# Control of nucleation and crystal growth of a silicate apatitic phase in a glassy matrix

D. de Ligny[(1)], D. Caurant[(1)], I. Bardez[(1,2)], J.L. Dussossoy[(2)], P. Loiseau[(1)], D.R. Neuville[(3)]

[(1)]*Laboratoire de Chimie Appliquée de l'Etat Solide, ENSCP, 11 rue P. et M. Curie, 75231 Paris cedex 05, France, dominique-deligny@enscp.jussieu.fr*
[(2)]*CEA/DEN/DIEC/SCDV, 30207 Bagnols-sur-Cèze, France, jean-luc.dussossoy@cea.fr*
[(3)]*Laboratoire de Physique des minéraux et des magmas, IPGP, CNRS UMR7047, boite 89, 4 place Jussieu, 75252 Paris cedex 05, neuville@ipgp.jussieu.fr*

***Abstract*** – *Nucleation and growth of crystal in an oxide glass was studied in a Si B Al Zr Nd Ca Na O system. The nucleation and growth process were monitored by thermal analysis and isothermal experiments. The effect of the network modifier was studied. Therefore for a Ca rich sample the crystallization is homogeneous in the bulk showing a slow increase of crystallinity as temperature increases. On the other hand, a Na rich sample undergoes several crystallization processes in the bulk or from the surface, leading to bigger crystals. The activation energy of the viscous flow and the glass transition are of same magnitude when that of crystallization is a lot smaller. Early diffusion of element is done with a mechanism different than the configurational rearrangements of the liquid sate. The global density and small size of the crystals within the Ca rich matrix confirmed that it would be a profitable waste form for minor actinides.*

## INTRODUCTION

The containment of long-lived separated radionuclides, minor actinides, would be profitably realized in a silicate apatite glass-ceramic obtained by controlled crystallization of parent glass. Indeed, the advantages of this waste form are numerous : glasses are easily handled and processed, both the glass and apatite can be used to immobilize actinides and the chemical fluctuations of the waste are absorbed in the glassy matrix [1]. Moreover apatite type compounds have shown their ability to incorporate Pu, minor actinide as Np, Am and Cm or lanthanide as Gd in a large range of composition. [2]

However, to be of real interest such an apatite glass-ceramic needs to be internally stress less and to present a good partition coefficient of the actinides between crystals and residual glass. These two last criteria meet if the formed crystals are in the bulk, highly concentrated and small in size. Therefore the ability to control the crystal nucleation and growth from the parent glass is crucial.

In this work, trivalent minor actinides were simulated introducing $Nd_2O_3$ in the glass composition. Intense apatite crystallization is obtained in the bulk for an alumino borosilicate glass containing up to 15 wt.% of $Nd_2O_3$. Electron microscopy, X-ray diffraction (XRD) and differential thermal analysis are used to characterize the devitrification.

The description of the nucleation and growth mechanism is done in term of activation energy. An important physical property is the atomic diffusion inside the glass matrix; it is often related with the atomic reorganization controlling the viscous flow. From viscosity measurement activation energy was determined and compared to the crystallization one.

As to evaluate the effect of local chemistry, two glasses are studied, a glass containing Na as network modifier and a second glass containing a mixture of Na and Ca (see TABLE I).

## EXPERIMENTAL

Three glasses were prepared from reagent grade oxide ($SiO_2$, $Al_2O_3$, $ZrO_2$, $Nd_2O_3$), carbonate ($Na_2CO_3$, $CaCO_3$) and boric acid ($H_3BO_3$) powders. All glasses were melted in platinum crucibles at 1400°C. The chemical homogeneity was achieved by regrinding at least once each sample after a first quench. Microprobe analysis and wet chemistry analysis confirm the nominal composition given in Table I. The Res.gl. sample is a residual end member of Ca.Na.gl. assuming that all the Nd was captured in a silicate apatite, $CaNd_8(SiO_4)_6O_2$ as observed in a previous work [3]. The Na.gl. was obtained by substituting in a molar basis all the Ca by Na from the Ca.Na.gl glass.
Samples prepared for isochronal recrystallization and viscosity measurement were annealed at 500°C for 5 days to relief internal strain and allow cutting for sample preparation. Being well below the Tg, around 600°C, this thermal treatment is taken as ineffective on the nucleation process. No further thermal treatment was done before experiment.

| TABLE I. Samples Nominal Composition in mol %. | | | |
|---|---|---|---|
| Oxydes | Ca.Na.gl. | Na.gl. | Res.gl. |
| $SiO_2$ | 53.7 | 53.7 | 54.2 |
| $B_2O_3$ | 9.3 | 9.3 | 10.5 |
| $Al_2O_3$ | 9.6 | 9.6 | 10.8 |
| $Na_2O$ | 15.1 | 21.7 | 17.1 |
| CaO | 6.6 | 0.0 | 5.1 |
| $ZrO_2$ | 2.0 | 2.0 | 2.3 |
| $Nd_2O_3$ | 3.7 | 3.7 | 0.0 |

**DIFFERENTIAL THERMAL ANALYSIS**

Two main observations can be done from thermal analysis. Grain size effect on the crystallization temperature characterizes a surface nucleation. The evolution of a temperature with the heating rate can always be interpreted in term of activation energy [4].

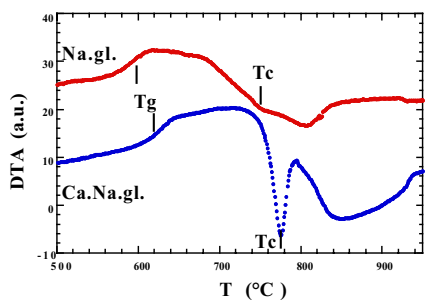

Fig. 1. Thermogram of the two glasses for a 10°C/mn. heating rate on a 20-40 µ fraction. Tg and Tc show respectively the glass transition and the maximum of crystallization.

As seen in Fig. 1, the two glasses undergo a two-stage crystallization above the glass transition. For Ca.Na.gl., a first exothermal peak is observed followed by a large and diffuse exothermal event. For Na.gl. the two events are closer and more difficult to identify.

**Grain size effect**

In the case of Ca.Na.gl., modifications of the size of the sample did not modify the position of Tc. The more diffuse effect at higher temperature however showed various shapes not very reproducible. It is deduced that Ca.Na.gl. undergoes a crystallization in the bulk, the sluggish variations at higher temperature would need further studies to determine if they are due to the sample or thermal diffusion effects as the powder melt and move inside the crucible.
In the case of Na.gl. at the opposite, the secondary peak was constant with grain size, when the first peak, notified Tc, increases and overlaps the constant exothermic effect. This last variation shows that a surface crystal growth mechanism coexists with growth in the bulk for Na.gl.

**Heating rate effect**

On a constant size fraction, 80-125 µm, the samples were heated up at rate from 1°C/mn to 55°C/mn. The temperature of crystallization was taken at the maximum of the first exothermic peak. The glass transition temperature of Na.gl. was noticeable at all rates, corresponding with a strong endothermic effect. It was not possible to trace Ca.Na.gl. Tg which is not surprising since heat capacities of Ca containing liquids are a lot smaller than Na ones.

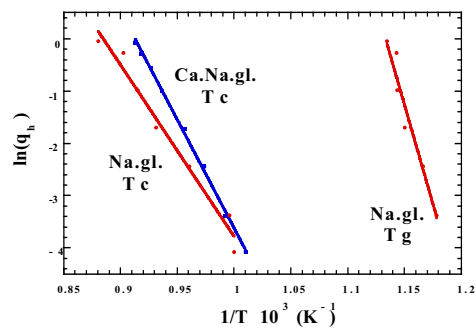

Fig. 2. Dependence of the temperature of crystallization Tc and of the glass transition temperature Tg versus heating rate.

The equation (1) was used as well for the temperature of crystallization and the glass transition. $q_h$ represents the heating rate.

$$\frac{d \ln q_h}{d(1/T_x)} = -\frac{\Delta H_x^*}{R}. \qquad (1)$$

The energy of activation $\Delta H^*_c$ is 345±7 kJ/mol for Ca.Na.gl. and 273±19 kJ/mol for Na.gl. For the glass transition only for Na.gl., $\Delta H^*_g$ is 630±70 kJ/mol. This strong contrast between activation energy of crystallization and of the glass transition shows that atomic reorganization around the growing crystal is a lot easier than the change of configurational states freed inside the liquid.

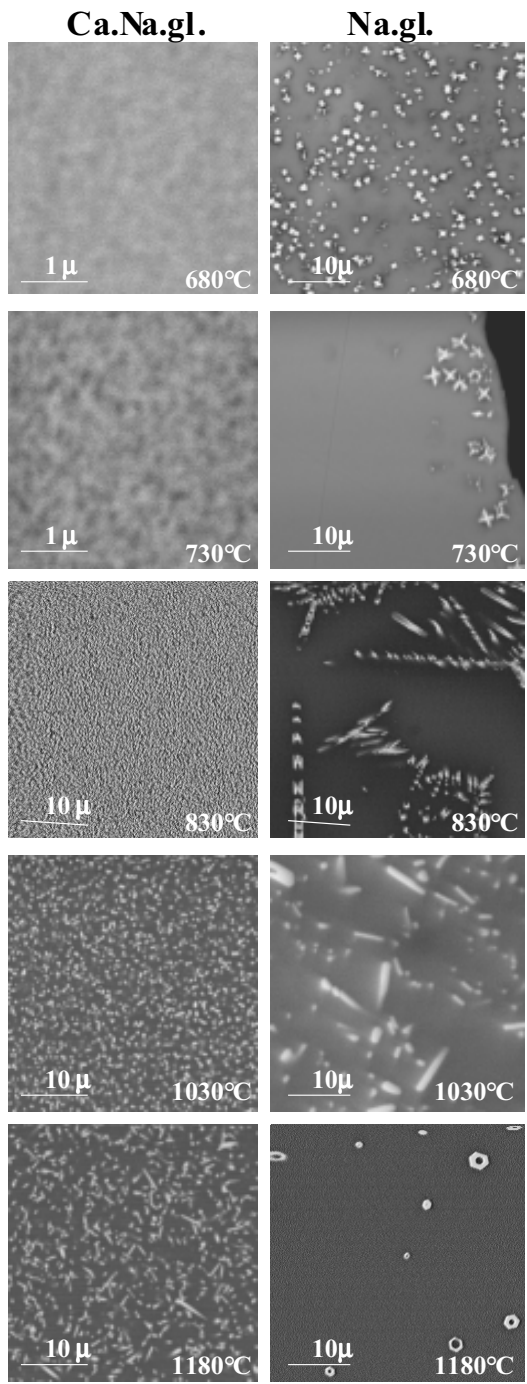

Fig. 3. Backscattered electron images obtained on samples heated at the notified temperature for 2 hours. Pictures on the left column are from Ca.Na.gl., those on the right column from Na.gl.

**SCANNING ELECTRON MICROSCOPE**

Isochronal preparation of 2h at different temperatures was done on glass pieces of 2g. The sample without initial nucleation stage was brought directly at high temperature in a tubular furnace. A thermocouple near the sample showed that the final temperature was reached within 4 minutes. At the end of the two hours, the sample was quenched at room temperature, part of it ground for X-ray diffraction and another mounted and polished for S.E.M. observations. Since the crystals we wanted to observe are rich in Nd which has a strong Z compared to the residual glass, backcscattered electron observation gives the best contrast.

The pictures, obtained and reported on Fig. 3, show a very different behavior for the two glasses. At low temperature, Ca.Na.gl. shows very fine granular structure under the resolution of our microscope. As temperature raises, the granular structure got sharper indicating a better repartition between rich and poor Nd phase. It is only above 1030°C that the silico apatite crystals could be seen still at a sub micron size. At higher temperature, as densification goes on, fine elongated shapes appear.

Na.gl. shows a more complex behavior. Isothermal heating at low temperature (680°C) leads to devitrification in the bulk of dense round crystallites where just 50°C higher, only small surface crystals could be observed. Large dendritic apatite crystals develops through out the sample from the surface at 830°C. At the highest temperature well crystallized apatite could be seen in the bulk. Depending on temperature, distinct growth mechanism took place in Na.gl. with nucleation on the bulk or on the surface. This variability of behavior of the Na rich glass seems to demonstrate the Na, as a strong network modifier, homogenizes a lot more the matrix and allows large cation mobility. All the X-ray patterns showed the presence of a single silicate apatite phase. However Na or Ca Nd-apatite has mostly the same patterns and cannot be identified in a quick study.
For Na.gl. the crystals were big enough to allow microprobe analysis that showed the stochiometric composition of $NaNd_9(SiO_4)_6O_2$. The microprobe analysis of the residual glass for the sample prepared at 1180°C showed that one third of the Nd was jailed inside the crystal phase. No such measures was possible on

Ca.Na.gl. due to the too small size of the crystals. However previous study showed the presence of a Ca apatite.

**VISCOSITY MEASUREMENTS**

Viscosity measurement was realized at low temperature below the crystallization region. The effect of temperature on viscosity is related only to the structural relaxation and is not perturbed by nucleation. The measurements were made with a creep apparatus, whereby the rate of deformation of the sample is measured as a function of an applied constant stress at a fixed temperature. [5]

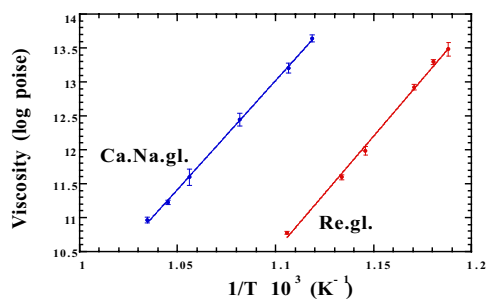

Fig. 4. Dependence of the viscosity with temperature.

Using the simple and well-know Arrhenius equation (2), the energy activation of the structural relaxation, $\Delta H_\eta^*$ can be estimated from the viscosity $\eta$.

$$\frac{d \ln \eta}{d(1/T)} = \frac{\Delta H_\eta^*}{R}. \qquad (2)$$

The value are 649±19 kJ/mol for Res.gl. and 615±19 kJ/mol for Ca.Na.gl. The nature of the internal rearrangement is similar for the two glasses. However as it can be seen the Res.gl. has a smaller viscosity. It can be foreseen that, as the silico apatite crystal appeared, atomic mobility increased in the residual glass, which help the growth mechanism.

If this activation energy is compared to those find earlier we can notice that the Na.gl. glass transition energetic (629±75 kJ/mol) is similar leading that structural relaxation and viscous flow have similar origin. The activation energies deduced from crystallization however are much lower, so they are underlined by an atomic diffusion that is not limited by the viscosity.

**CONCLUSIONS AND PERSPECTIVES**

If nucleation and growth mechanisms appear to be simple in the case of the Ca rich glass, the Na glass showed several behaviors. Calcium holds a highly clustered environment helping crystal nucleation. This Ca function is in agreement in the higher crystallization activation energy for Ca.Na.gl. The ratio Ca/Na is critical to control nucleation process.
The poorly crystallized Na.gl. leaded to a segregation of one third of the Nd inside the crystal phase, it can be expected a better result for Ca.Na.gl.
Since shape of the crystals, small and dense, and chemical separation are promising, this system is confirmed to be a profitable waste form for minor actinides.